# Derivation of Principle of Extreme Physical Information

B. Roy Frieden,[1] Robert A. Gatenby[2]
[1]*College of Optical Sciences, University of Arizona, Tucson, Arizona 85721*
[2]*Departments of Radiology and Integrated Mathematical Oncology, Moffitt Cancer Center, Tampa, Florida 33612*

**TABLE OF CONTENTS**

## 1. ABSTRACT

Let data from a natural effect (physical, chemical, biological, etc.) obey an unknown probability density function $p \equiv q^2$, $p \equiv p(x), q \equiv q(x)$, $x$ a scalar (for now) coordinate such as a particle position. Thus $q(x)$ is an unknown amplitude function. It may be found using the principle of Extreme physical information (EPI). The EPI principle is straightforwardly derived as follows. The Fisher information level $I$ in data from the unknown statistical source effect is defined as (i) $I \equiv 4 \int dx\, {q'}^2$ where $q' \equiv dq/dx$. It was shown in a previous publication that, owing to L. Hardy's 5 mathematical axioms defining all known physics, $I = max$. Therefore, its first variation (ii) $\delta I = 0$. The total effect is modeled as an information channel $J \rightarrow I$, with $J$ the information level of the *source* effect giving rise to *data* of information level $I$. The essential difference between informations $I$ and $J$ is that $I$ is generic, obeying (i) for all source effects, whereas $J$ is specific to the particular source at hand. Hence, $J$ does *not* have the integral form (i) but, rather, a general form (iii) $J \equiv \int dx\, j[q(x), s(x)]$, with $j$ some continuous function of its arguments and $s(x)$ some known physical source such as mass, current, biological fitness, etc. Fisher information has the intrinsic property of decreasing under irreversible coarse graining of the system, such as measurement. Therefore the data information obeys $I \leq J$. This is equivalently a statement $I = \kappa J$, where $0 \leq \kappa \leq 1$. Then variation $\delta I = \kappa \delta J$ so that property (ii) gives (iv) $\delta J = 0$ as well. Then combining (ii) and (iv), $\delta(I - J) = 0$. Or, $I - J = extremum$. What is the nature of the extremum? Let $\mathcal{L}$ be the integrand of the functional $I - J$. Eqs. (i) and (iii) give (v) $\mathcal{L} = 4q'^2 - j[q(x), s(x)]$. The Legendre condition states that the extremum is a minimum if $(\partial^2 \mathcal{L})/(\partial q'^2) \geq 0$. Note from (v) that $q'$ exists only in the first term of $\mathcal{L}$. Then $(\partial^2 \mathcal{L})/(\partial q'^2) = +8$. Therefore by the Legendre condition the statistical source effect obeys a principle (vi) $I - J = minimum$, the EPI principle. This is a knowledge-based approach to natural law, where the knowledge $j$ is some aspect of the unknown source effect, not necessarily its energy or action values. This accounts for its past use to derive phenomena of both physics and other sciences, including population biology, cancer growth, chemistry and econophysics. Given its wide application, a derivation of EPI was long sought, and has now been found.

## 2. LEAST ACTION

A major accomplishment of the Age of Enlightenment was discovery of the principle of 'least action.' This is, in Lagrangian form,

$$\mathcal{L} \equiv T - V = minimum. \tag{1}$$

Least action forms have the virtue of giving correct physical laws as their outputs. (Sometimes 'maximum action' is the case.) For example, in classical mechanics $T$ is the kinetic- and $V$ the potential energy, and the least action principle gives rise to Newton's laws of motion. This principle is based upon a completely phenomenological view of physics. For example, both $T$ and $V$ are energy-dependent quantities.

### 2.1 On generalizations

As found by H. von Helmholtz and others,[1] there are appropriate action forms $T - V$ for all fields of physics. From this point of view, a principle of least action is behind all physics; and consequently, it is *the most basic physical effect* of all. However, as a limitation to its use, least action requires that quantities $T$ and $V$ be known energies or energy-dependent terms.

### 2.2 Limitations of use

But not all natural effects can be specified by energy-dependent quantities $T$ and $V$. Examples are particle population dynamics, biological laws of allometric growth, and most aspects of econophysics. Is there, then, an alternative variational approach that does not require knowledge of energies?

### 2.3 Alternative of Information-based approach

In fact, as seen below, effects such as particle population dynamics, biological laws of allometric growth and aspects of econophysics can be derived using quantities $T$ and $V$ that are Fisher informations $I$ and $J$, respectively. In contrast with the phenomenological point of view of the least action approach, this information approach is *epistemological* in nature.  It describes the unknown effect by its ability to convey knowledge, or, specifically *information*, about its nature.

This epistemological viewpoint has demonstrated that least action is *not unique* in allowing the derivation of physical effects. For example, over the past 20 years the concept of *maximum* Fisher *information*[2-8] and its use in the principle of Extreme physical information (EPI)[5,9-17] have likewise served this purpose, as well as deriving laws of biology, chemistry and econophysics. This approach was anticipated by the many scientists that, in the past, had sought a principle that would derive non-physical effects, such as biological population growth[5,16] and economic theory,[9] as well as physical effects.

## 2.4 Question of prior rationale for least action

Aside from so voicing the need for freeing the analyst from the need for knowing system energy terms $T$ and $V$, the question of *why*, a priori, 'least action' works to derive physics, was long ago voiced. No less a scientist than Schrodinger, e.g., called it "incomprehensible"[18] why the principle $T - V = minimum$ works to derive his wave equation.  It seems that the $T - V$ form for 'action' has no known prior *physical* significance other than being a device to derive physical laws by 'reverse engineering' them.  Why not instead, e.g., a principle $T + V = minimum$ or even $T/V = minimum$? Indeed, why only energy terms in the first place? True, the Planck constant *h* has the units (energy-time, or momentum-position) of action, but that doesn't account for these uses of the particular form $T - V$.

# 3. J. WHEELER'S VIEW

An essential part of nature is its dynamics. Understanding this is based, ultimately, upon observing them.  In fact J. Wheeler emphasized[19] that such observation is actually *formative* of its physical expression. But, in fact, how well can we observe, even with perfect detectors?  Moreover, insofar as knowledge ultimately traces to the measurement of physical parameters, how well can we *know*? This is an epistemological question that turns out to have important physical consequences.

Consider that, even with perfect detectors, we can only sense reality imperfectly, i.e. as *projections* of it (think C.T. scan images). This provides some guidance as to how physics and *knowledge of it* interrelate. Our *inability* to perfectly sense reality traces at least as far back as "the cave" parable of the philosopher Plato. By this parable, people who were constrained to live their lives within a cave could only gain knowledge of objects existing outside by observing the shadows they cast on the cave walls.  The knowledge so gained was but a 'shadow' of reality, i.e. very imperfect. He believed that absolute perfection does exist, in the form of the outside objects. But that perfect *knowledge of* reality by observers (human or not) is not possible, since all observations are necessarily imperfect.

As mentioned above, recognizing and quantifying these epistemological limitations has led to the derivation[2-16] of many known (and some previously unknown) statistical laws of nature.

# 4. DATA INFORMATION *I* OF FISHER

The approach is grounded in classical information theory, invented largely by R.A. Fisher[20] circa

1920.  (Note: This is not R.A. Shannon's information, invented for other purposes in about 1945.) A system is characterized by a probability law $p(x)$ on random fluctuations $x$ from a state parameter *a.* Its value is to be found. The fluctuations might be due to random diffusion. Thus, even with perfect detection, any observation is not the ideal value $a$ but, rather, a datum $a + x$ with $x$ a random sample from the system law $p(x)$.  Thus, to sensibly estimate $a$, the system is observed a number of times, and an estimate of $a$ is formed as some chosen function of the data (say, their average).

Let the law $p(x)$ have an amplitude $q = \sqrt{p}$, $q = q(x)$. Note that $q$ has, in general, the role of a probability amplitude, as in quantum theory.

## 4.1 Definition

The concept of the Fisher information about the state $a$ in a single observation is utilized, as below, to understand how accurately the value of $a$ can be known from a single measurement.  The information is most conveniently represented in terms of amplitude $q(x)$, rather than $p(x)$, as

$$I = 4 \int dx \left(\frac{dq}{dx}\right)^2, \; q = q(x). \tag{2}$$

This assumes a property of shift invariance for the law $q(x)$. Also, the integration is taken over the entire range of the random variable *x*. Note that, in Eq. (2) for a quantum effect $q(x) = \psi(x)$, the wave function, and the integrand $(dq/dx)^2$ is replaced by $\nabla \psi^* \cdot \nabla \psi$ where $\nabla$ is the gradient operator and $*$ denotes the complex conjugate.

By K. Popper's criterion of negation, any effect that is claimed to be physical (as opposed to, say, metaphysical) must make predictions that can be falsified.  To falsify the effect requires, at the very least, its accurate observation in a well-defined state *a*. (Note that this is the very measurement scenario addressed by L. Hardy's axioms mentioned later.)

The above integral for *I* describes the information in a single observation of *any physical effect* defined by a shift-invariant law $q(x)$ or $p(x)$.  For example in measuring the ideal position *a* of a quantum particle *x* is the fluctuation error from *a* and The expression for $I$ is easily generalized to multiple observations by the usual replacements $x \to \mathbf{x}, dx \to d\mathbf{x}$ in the integrand.

## 4.2 Generic nature of *I*

With these replacements as needed, $I$ has the same integral form (above) for all effects. This is because *it regards all data as generic*, leaving to the source information *J* (as in Plato's "cave parable") the role of defining the *specific* effect. Information *J* is quantified at Eq. (4) below.

Since the 1920's the above integral for $I$ has been mainly used to define how well the state $a$ can be estimated (see below).  However, this is mere use as a diagnostic. Moreover, our ultimate aim is not to merely estimate the state $a$ but, rather, *to estimate the system*: the form of either $q(x)$ or $p(x)$.  The integral $I$ will likewise prove indispensable for *this* purpose. At this point, we emphasize that the above integral form for $I$ is, in particular, not generally dependent upon prior knowledge of *system energy*.  Thus, neither will formation of the estimate of $q(x)$ or $p(x)$.

## 5. OBSERVING IS A LOSSY PROCESS

Consider the measurement of a state parameter $a$, say a particle mass. The "Platonic ideal" is that, generally speaking, perfection exists but that man seldom observes it. In fact, according to modern measurement theory, any macroscopic measurement of a parameter is *intrinsically* lossy, i.e. irreversible in nature. In order to produce a definite reading of the state $a$ of the system, the detector in use interacts with the system. It is this very interaction, causing an irreversible exchange of information and energy with those of the system, that gives rise to the output measurement. However its irreversible nature amounts to a 'coarse graining'[5,21] process. This, by definition, causes the observed data to suffer *a loss* of Fisher information

$$\delta I \leq 0 \qquad (3)$$

from its intrinsic value $J$ prior to measurement. Thus, with $I = J + \delta I$ it must be that

$$I \leq J. \qquad (4)$$

# 6. SOURCE INFORMATION $J$

By Plato's parable, a given system has a perfectly well-defined state of a property $a$, but this state cannot be known exactly. Thus, state $a$ is perfectly described by some finite, fixed level $J$ of information about parameter $a$. However, any data information level $I$ that is less than $J$ could only imperfectly define state $a$. Thus, inequality (4) in effect proves a modern version of Plato's parable:

Coarse graining demarks the transition from a quantum to classical universe[22,23]; (corresponding to the transition from efficiency $\kappa = 1$ to $\kappa = \frac{1}{2}$ as discussed in Sec. 6.4.)

## 6.1 Examples of $J$

In providing a complete description of system state $a$, information $J$ is intrinsic to the observed system. It exists at the source effect and, therefore, must be expressible in terms of its defining physical properties. For example, in describing: (a) quantum observation of the position of a particle,[5,9] $J$ is proportional to the square of the particle's mass; (b) cell growth, $J$ increases with reproductive fitness[5,7,16]; (c) the growth of investment capital in econophysics, $J$ increases as the expected value of the production function[8]; (d) cancer growth, $J$ increases with cancer mass;[5,11] (e) the growth of competing populations, $J$ is proportional to the mean-squared fitness over the populations.[5,16] Notice that in no application (b)-(e) is there an explicit dependence upon energy – kinetic or potential – or upon their difference, the 'action.'

## 6.2 Physics as transition from substance to observation; from being to becoming

In contrast with the intrinsic or source information $J$, the observer collects a level of information $I$ *about* $a$ in *data* collected *from* the system. He/she is at the receiving end of a basic flow of information

$$J \rightarrow I, \text{ or } being \rightarrow becoming. \qquad (5)$$

The 'being' defines the ideal (as we will see, maximum) level $J$ of Fisher information in the source effect in terms of physical sources. The direction of the arrows in (5) also indicates that cause precedes effect: the physical source gives rise to the data, and not the other way around.

Information $J$ is taken to obey the general form

$$J \equiv \int dx \, j[q(x), s(x)]. \tag{6}$$

Here $j$ some continuous function of its arguments defining a source effect. In particular, $s$ defines the physical source, such as mass, current, biological fitness, etc. (in particular, not necessarily energy). Note also that $J$ does not depend explicitly upon the gradient $q'(x)$. Such explicit dependence is reserved for information $I$, as defined in the generic form (2). Also, if $J$ depended upon $q'(x)$ there would no longer be a definite distinction between 'being' and 'becoming' in the flow Eq. (5).

As an example, the flow (5) of information occurs during its transfer (say, via photons in a microscope) from a physical source (in the microscope slide) at information level $J$ and in state $a$, to the observed level $I$ in data space (on the observer's retina).

Thus, although $J$ is an information, it is expressed in terms of physical properties of the system. As we noted, this ideal system then 'becomes' observed as data of information level $I$. In the book *Parminedes*, by Plato, the Greek philosopher Parmenides noted that this process defines an ontology of ongoing human activity, and that it is *generally imperfect.* The latter has been proven at Eq. (4) in the form of a general reduction in the level of observed information.

## 7. EPI DERIVED FROM L. HARDY'S MATHEMATICAL AXIOMS

The mathematician Lucien Hardy discovered[24] a system of five mathematical axioms that are the basis for all known physics, both classical and quantum. For our purposes, chief among them is the axiom that in one observation of a system the number $N_0$ of distinguishable states is a maximum value. From this it was deduced[2] that the system has maximum Fisher information,

$$I = maximum. \tag{7}$$

(See a simplified proof in the Appendix.) The principle (7) has been used in many applications.[3,6,7,8,17]

### 7.1 Practical need for constraints

However, note that the use of Eq. (7) *alone* as a principle of estimation would be useless: The integral form $4\int dx \, (dq/dx)^2$ for $I$ can approach infinite value if some system gradient values $dq/dx$ are sufficiently large in absolute value. Therefore, setting $I = maximum$ subject to no constraint term $J$ would always give $I = \infty$, a useless principle for finding $q$. By comparison, the addition to $I$ of an appropriate constraint term such as $-J$ (as taken up below) acts to keep gradient sizes $dq/dx$ under control. The constraint per se arises out of some level of prior physical knowledge of the unknown probability law (as in Ref. 3 where the mean kinetic energy is known, or that it obeys continuity of flow as in the example below).

### 7.2 Verifying Plato

What does Eq. (7) say about source information $J$? By inequality (4) $J$ = maximum as well, in fact a larger maximum than $I$. This verifies Plato's thesis that perfection only exists at the source. By comparison, its observation is generally less than perfect, as previously discussed (see end of Sec. 5).

### 7.3 Derivation of EPI

We are now in a position to derive the EPI principle $I - J = minimum$. Let $\mathcal{L}$ be the integrand of the functional $I - J$. From Eqs. (2) and (6),

$$\mathcal{L} = 4q'^2 - j[q(x), s(x)]. \tag{8}$$

The *Legendre condition* states that the extremum is a minimum if $(\partial^2 \mathcal{L})/(\partial q'^2) \geq 0$. Note from Eq. (8) that $q'$ exists only in the first term of $\mathcal{L}$. Then directly $(\partial^2 \mathcal{L})/(\partial q'^2) = +8 > 0$. Therefore by the Legendre condition the statistical source effect obeys a principle $I - J = minimum$. Also, due to inequality (4), $I \leq J$, or equivalently, $I = kJ$ with $0 \leq k \leq 1$. Therefore in summary

$$I - J = \text{minimum}, \quad \text{where } I = \kappa J, \ 0 \leq \kappa \leq 1 \tag{9}$$

This is called the principle of Extreme physical information or EPI (as in 'EPIstemology'). QED

Principle (9) may be interpreted on two levels:

(1) **Philosophical level**

The EPI principle (5) establishes the transition of substance, or ideal *being*, into *becoming*, or observed data, as the route to understanding an unknown effect. These are represented by their respective information levels $J$ and $I$. Coefficient $\kappa = I/J$ measures the efficiency of the information transfer about $a$. That the information loss is a minimum serves *to quantify* Plato's thesis of such a loss.

Of course for purposes of gaining knowledge it would be best if the received information obeyed $I = J$, the source level. However, the 2$^{nd}$ Eq. (9) indicates that the efficiency coefficient $\kappa \leq 1$. As we saw, this followed from the *irreversible* nature of making a measurement.

(2) **Practical level**

EPI principle (9) also defines a variational problem whereby the minimum is found through variation of either unknown amplitude law $q(x)$ or probability law $p(x)$. This establishes the natural law governing the data. Note that EPI actually consists of two conditions (9). The second, $I = \kappa J$, has no counterpart in the least action approach (1). It gives practical advantage in cases $\kappa < 1$, in particular where $\kappa = 1/2$ defining all classical physics. There it allows unknown law $q(x)$ to be found as the simultaneous solution[5,17] to *the two* conditions (9).

A second practical benefit is that they provide an optimum estimate of the unknown state parameter $a$. The logic here is that if the parameter $a$ is to be best estimated this requires that the system 'behind it' be best estimated as well. Nature evidently follows this logic, since the EPI principle works[3,5,9-17] and the best estimate (in the sense of minimum mean-square error) of $a$ is often achievable.[5,6,11] However, even the best estimate is not perfect. By the Cramer-Rao inequality (14) below, even in this ideal scenario $I = J$ where *I* is finite, there is still residual rms error *e* in the estimate of the state *a*. The Platonic ideal cannot be realized for the estimate of the parameter. But, even so, since $I$ is maximized the error *e* is *minimized*. Thus, reality cannot be perfectly known, but it can be known with minimal error.

Note that, in all the preceding we assume the use of a perfect detector, i.e. one that does not add fluctuations of its own to the readings. This enabled us to concentrate on the system, establishing its nature per se.

## 7.4 Verifying Wheeler's 'participatory universe'

As a mathematical possibility, the minimum attained by EPI principle (5) could be very negative: At its worst, conceivably $\kappa = 0$ and $I - J = -J$. Here the desired result $I \approx J$ would be far from satisfied. However, in all applications of principle (5) this solution has never occurred. The largest departure from the ideal result $I = J$ has been $I = (1/2)J$, representing a 50% loss of source information. This occurs in all scenarios of classical physics (mechanics, electromagnetism, etc.). By comparison, the ideal result $I = J$, where efficiency $\kappa = 1$ and there is zero loss of information, occurs in all quantum scenarios. This includes derivation[5] of the wave equations of Klein-Gordon, Schrodinger, Dirac, and Rarita-Schwinger, as well as the Einstein-Podolsky-Rosen equation of quantum entanglement. These results serve as examples of Wheeler's[19] view (mentioned above) that the observer participates in creation of the effect. Thus classical physics, which is ordinarily observed on a macroscopic (coarse) level, wastes 50% of the intrinsic information $J$; whereas quantum physics, ordinarily observed on a microscopic (fine) level, does not waste any of the intrinsic information.

In summary, Hardy's axioms establish three important statements of knowledge acquisition:

(1) The EPI principle (9) that $I - J = minimum$.

(2) The principle (7) that $I = maximum$ when supplemented with constraint equalities.

(3) $J$ is the maximum, and therefore intrinsic, information level of the source. This proves the central theme of Plato's parable that the information $J$ at the source (the person outside the cave casting the shadow) represents maximum, or absolute, truth.

## 8. ILLUSTRATIVE EXAMPLES

The idea behind EPI is best shown by elementary examples. We next use it to derive both the electromagnetic wave equation[5] and the Heisenberg uncertainty principle.[4,5]

### 8.1 Electromagnetic wave equation

Here the primary source effect is electromagnetic wave motion. Our primary aim is to find the law obeyed by the four-vector amplitude law $\mathbf{q}(\mathbf{r}, t)$, $\mathbf{q} \equiv (q_1, \ldots, q_4)$, $\mathbf{r} = (x_1, x_2, x_3)$ describing it. That law will be the electromagnetic wave equation. And from it, Maxwell's equations follow.

In any EPI problem, some prior physical knowledge must be at hand so as to determine the intrinsic source information $J$. Once determined it is used to simultaneously solve the two EPI conditions (4) for the same law $\mathbf{q}(\mathbf{r}, t)$. Finally, $\mathbf{q}(\mathbf{r}, t)$ is identified with the electromagnetic vector potential. The concept of energy is not explicitly used.

In order to find $J$ an input effect must be known that is intrinsic to the effect at hand, here that of electromagnetism. Here the chosen effect is the four-dimensional (covariant) Lorentz condition

$$\frac{1}{c}\frac{\partial q_4}{\partial t} + \sum_{n=1}^{3} \frac{\partial q_n}{\partial x_n} = 0. \tag{10}$$

Quantity $c$ is the speed of light in vacuum. Eq. (6) expresses continuity of flow of amplitudes $q_n$ in space and time.

The use of condition (10) gives[5] as the intrinsic information

$$J = 4c \iint d\mathbf{r} dt \sum_{n=1}^{4} E_n J_n, \quad J_n = J_n(\mathbf{q}, \mathbf{j}, \rho). \tag{11}$$

Here **j** is the current density (three directional components) and $\rho$ is the charge density. The $E_n$ are constants to be found that weight information components $J_n(\mathbf{q}, \mathbf{j}, \rho)$. The latter are found below.

We emphasize at this point that information $J$ is, in particular, *not the electromagnetic energy*. The EPI approach to this problem does not depend upon knowledge of its energy, in contrast with the corresponding least-action approach.

The minimization in EPI is accomplished by working with a $Lagrangian$ $\mathcal{L} = i - j$, with $i, j$ the respective integrands of $I$ and $J$. From (2) the information Lagrangian $i = 4c \sum_{n=1}^{4} \nabla q_n \cdot \nabla q_n$, and $j$ is the integrand of Eq. (11).

In general the solution $q_n, n = 1, ..., 4$ to a continuous EPI problem obeys the Euler-Lagrange differential equations

$$\sum_{k=1}^{3} \frac{d}{dx_k}\left(\frac{\partial \mathcal{L}}{\partial q_{nk}}\right) + \frac{d}{dt}\left(\frac{\partial \mathcal{L}}{\partial q_{n4}}\right) = \frac{\partial \mathcal{L}}{\partial q_n}, \text{ where } q_{nk} \equiv \frac{\partial q_n}{\partial x_k}, k = 1,2,3, \text{ and } q_{n4} \equiv \frac{\partial q_n}{\partial t}. \quad (12)$$

The solution here[5] obeying condition (10) is the wave equation

$$\Box \mathbf{q} = B \mathbf{J_S}, \ \Box \equiv \frac{1}{c^2}\frac{\partial^2}{\partial t^2} - \nabla^2, \text{ and } \mathbf{J_S} \equiv (\mathbf{j}, c\rho). \quad (13)$$

$B$ = constant. Also, efficiency constant $\kappa = 1/2$. Or, 50% of the information is lost during observation of the electromagnetic effect.

By the Born approximation, probability amplitudes are effectively potentials for single-particle exchange forces. Thus the four probability amplitudes $\mathbf{q} = \mathbf{A}$, the four-potential of electromagnetic theory. With these identifications, (13) becomes the electromagnetic wave equation for the four-potential. As is well known, Maxwell's equations follow directly from the wave equation.

## 8.2 Heisenberg uncertainty principle

Assume, for this problem, position and momentum fluctuations $x$, μ about mean values of zero. The Cramer-Rao inequality [2,5,6,11]

$$e^2 \geq 1/I \quad (14)$$

shows how the mean-squared error $e^2$ in estimating the system state *a* depends upon the level of Fisher information $I$ in the data. As expected, the higher the information the lower the error.

We consider the problem of measuring the position *a* of a particle of mass *m* moving with constant energy $W$. In that problem, the source information $J = b<W - V(x)> = b<E_{kin}>$ for a known constant $b$. Of course the mean kinetic energy also obeys $<E_{kin}> = <\mu^2>/2m$ with μ the momentum. Also, in quantum applications[5,10,12,15] of EPI information, as discussed above $I = J$. Then $I = b<E_{kin}> = b<\mu^2>/2m$. Then the above C-R inequality gives $e^2 \geq \beta^2/<\mu^2>$, or

$$e^2 <\mu^2> \geq \beta^2 \quad (15)$$

with the constant $\beta = \hbar/2$. Since this is a time-stationary problem, both uncertainties $e^2$ and $<\mu^2>$ occur simultaneously, so it shows the usual reciprocity in their uncertainties. If momentum were instead measured, the same uncertainty relation would result.

# 9. SUMMARY OF ATTRIBUTES OF EPI APPROACH

This overall information effect $I - J = minimum$, where $J = maximum$ and $I \approx J$, states that nature is, in a sense, 'kind' to observers. What is observed tends to be correct. Among other things this allows the observer to effectively find sources of nutrition, desired mates for purposes of reproduction, etc. It is thus consistent with the so-called 'strong anthropic principle.' This assumes a universe whose constants happen to accommodate cosmological and biological evolution as we know it. Simply put, in the absence of these constants we wouldn't be here. The existence of such a universe is, in turn, consistent with the existence of a multiverse consisting of universes with all possible combinations of universal constants. Thus, nature is very helpful to seekers of knowledge about *this* world at least. As we have seen, this is as follows:

1. The information *J* that is intrinsic to natural effects is *maximal* (i.e. maximum, subject to a constraint).

2. That $I \approx J$ is a statement that the information collected about state *a* in *a single* observation reflects the maximum information level that nature provides. Thus, what you see is, most likely, what is actually there, plus or minus modest error. This is further quantified, as follows, in the multiple-observation case.

3. Consider *repeated* observations of a parameter $a$. The accuracy of any estimate of $a$ based on many *observations* obeys the above Cramer-Rao inequality.[2,4,5] This states that the smallest possible root mean-square (rms) error *e varies inversely with the level of information I*. This smallest error is attained when an 'efficient' estimator function[5,10] of the data is formed (e.g. the simple arithmetic mean is efficient, when fluctuations *x* from *a* are Gaussian). But, as we found above, any observation carries information $I \approx J =$ *maximum* amount possible. Then the error $e \propto 1/\sqrt{I}$ is *further minimized*.

4. The EPI principle of minimization $I - J = minimum$ also constitutes a variational principle that gives *correct solutions* for the unknown phenomena $q(x)$ or $p(x)$. Moreover, applying the principle has advantages over using, say, least action. Whereas the latter requires knowledge of the system energies, such as *T* and *V*, the use of EPI does not. This is fortunate, since *non-energy* based Lagrangians have been found necessary for deriving laws of (i) cancer growth,[5,11] (ii) biological allometry,[5,17] (iii) cosmological allometry,[17] (iv) the de Broglie wave 'hypothesis' (no longer a mere hypothesis),[15] (v) thermodynamics using Fisher information in place of the entropy,[6] (vi) the Euler equation[12] of chemical density functional, laws of economic investment,[9] and of population dynamics.[5,16]

5. Physics can be usually derived in two different, but mathematically parallel, ways: as physically dynamic processes (the historic, least action route) or as information relay processes (the Fisher EPI route). However, as was noted in Sec. 6.3(2), a vital mathematical advantage of the information approach is the second principle (9) in EPI, which has no counterpart in the least action approach. It allows the unknown $q(x)$ to be found as the simultaneous solution to the two EPI Eqs. (9).

6. An approach to deriving specifically non-physical effects, such as those of economics or biology, that would be as effective as that of least action was long sought.[25] EPI provides an answer.

7. The resemblance between the EPI and least action recognizes that physics is intrinsically *both* phenomenological and epistemological. In fact EPI *can*, under special conditions, *become the least action principle*. We started out asking what the action Lagrangian 'action' $T - V$ represents *on its own*: prior to its use in deriving laws of physics. From the preceding, it arises as the loss $I - J$ of *information* due to observation of the physical effect in question *when, in particular, I ∝ T and J ∝ V*. This occurs in the energy representations of classical- and quantum mechanics, where the information-based EPI principle *I – J* = minimum morphs into one of least action, $T - V = minimum$. Thus, under these conditions least action does have *prior significance*, albeit of *epistemological* origin.

## APPENDIX: WHY A PRIORI $I = Max$.

We give a simple, one-dimensional version of the proof given in Ref. 2 of why the Fisher information $I = max$. of a physical system. The reader is invited to view the full proof, which is derived much more generally. In brief, this follows from L. Hardy's mathematical axioms;[24] these are the basis for all currently known physics.

Let the system be of length $L$, and its parameter state $a$ is to be estimated from *a single* measurement of it. The measurement gives an imperfect data value $a + x$, where $x$ is a random sample from the system's probability law $p(x)$. How well can the parameter be estimated?

Taking the square root of Eq. (9) establishes that the root-mean-square (rms) error $e$ in any such estimate obeys

$$e \geq 1/\sqrt{I} \tag{A1}$$

with $I$ the Fisher information. So the larger the information $I$ the smaller will be the rms error. Reciprocating Eq. (A1) gives

$$1/e \leq \sqrt{I}. \tag{A2}$$

On this basis, can we compute the maximum possible number $N_0$ of distinguishable states $a$ possible for the system. This is the number resulting by packing them maximally close together, i.e. mutually spaced by error distances $e$. Since they cover the distance $L$

$$N_0 = L/e. \tag{A3}$$

But also, multiplying (A2) by $L$ gives

$$L/e \leq L\sqrt{I} \tag{A4}$$

Using this in (A3) gives

$$L\sqrt{I} \geq N_0. \tag{A5}$$

Now, the mathematician Lucien Hardy has discovered five axioms[24] that turn out to provide the foundation for all known physics. The chief one for our purposes is that

$$N_0 = max. \tag{A6}$$

That is, the number of distinguishable values of parameter $a$ that can be defined by a single measurement of it is a maximum.

Since length $L$ is fixed, Eqs. (A5) and (A6) together state that $\sqrt{I} \geq max.$, or $I \geq max.$, implying that

$$I = max. \tag{A7}$$

The system, regardless of its nature, must have a maximum number of distinguishable levels in a single observation of it.
QED

## CORRESPONDING AUTHOR INFORMATION
Tel: (520) 621-4904
Fax: (520) 621-3389
E-mail: roy.frieden@optics.arizona.edu

## RUNNING TITLE
Derivation of EPI